\documentclass[11pt]{article}
\usepackage{amsmath,amsfonts}
\def \beq  {\begin{equation}}
\def \eeq  {\end{equation}}
\def \beqar {\begin{eqnarray}}
\def \eeqar {\end{eqnarray}}
\def\sqr#1#2{{\vcenter{\vbox{\hrule height.#2pt
\hbox{\vrule width.#2pt height#1pt \kern#1pt
\vrule width.#2pt}\hrule height.#2pt}}}}

\def\la {{\langle}}
\def\ra {{\rangle}}

\def\bp {\bar p}

\def\bD {\bar{D}}
\def\bA {\bar{A}}

\def\bu {\bar{u}}

\def\del {\partial}
\def\bdel{\bar{\partial}}

\def\N {{\cal N}}

\def\half{\textstyle{1\over 2}}

\begin{document}
\fontfamily{cmr}\fontsize{11pt}{17.2pt}\selectfont
\def \CMP {{Commun. Math. Phys.}}
\def \PRL {{Phys. Rev. Lett.}}
\def \PL {{Phys. Lett.}}
\def \NPBProc {{Nucl. Phys. B (Proc. Suppl.)}}
\def \NP {{Nucl. Phys.}}
\def \RMP {{Rev. Mod. Phys.}}
\def \JGP {{J. Geom. Phys.}}
\def \CQG {{Class. Quant. Grav.}}
\def \MPL {{Mod. Phys. Lett.}}
\def \IJMP {{ Int. J. Mod. Phys.}}
\def \JHEP {{JHEP}}
\def \PR {{Phys. Rev.}}
\def \JMP {{J. Math. Phys.}}
\def \GRG{{Gen. Rel. Grav.}}
\begin{titlepage}
\null\vspace{-62pt} \pagestyle{empty}
\begin{center}
\rightline{CCNY-HEP-07/5}
\rightline{October 2007}
\vspace{1truein} {\Large\bfseries
A Note on Graviton Amplitudes for New Twistor}\\
~\\
{\Large\bfseries String Theories}\\
\vskip .1in
{\Large\bfseries ~}\\
\vspace{.5in}
{\large V. P. NAIR}\\
\vspace{.1in}{\itshape Physics Department\\
City College of the CUNY\\  
New York, NY 10031}\\
E-mail:
{\fontfamily{cmtt}\fontsize{11pt}{15pt}\selectfont vpn@sci.ccny.cuny.edu}

\vspace{.4in}
\centerline{\large\bf Abstract}
\end{center}
A new class of twistor string theories were recently constructed by
Abou-Zeid, Hull and Mason. The most interesting one of these theories has the particle content of
${\cal N}=8$ supergravity. Arguments are given for the vanishing of the
$(+ - -)$-helicity amplitude for gravitons which suggest that this theory describes a chiral
${\cal N}=8$ supergravity rather than Einstein supergravity.

\end{titlepage}
\pagestyle{plain} \setcounter{page}{2}
Recently, a new set of twistor string theories has been proposed which has the potential of describing a number of interesting field theories as twistor string theories \cite{AHM}.
This was done by gauging the Berkovits form of the Witten-Berkovits twistor string theory with respect to a holomorphic one-form on twistor space \cite{witten, berk}. The Witten-Berkovits string theory
describes 
$\N =4$ supersymmetric Yang-Mills theory coupled to $\N =4$ conformal supergravity \cite{witten-berk}.
There are two reasons why one might seek to modify this string theory.
If the twistor string theory is to be used for calculations in Yang-Mills theory, then it is necessary to decouple (or eliminate) the conformal supergravitons which can otherwise contribute in loops. Being conformally coupled, there is no dimensional parameter which can be tuned to eliminate them. Secondly, since conformal gravity is a theory with fourth-order derivatives, the linearized equations of motion have additional solutions which do not have a positive Hilbert space norm and hence leads to  loss of unitarity. If a new gauge symmetry can be implemented which eliminates these unphysical modes, then both these problems can be eliminated. Abou-Zeid, Hull and Mason introduced  a new gauge symmetry which has the potential for doing just this \cite{AHM}.
A number of different theories, which correspond to different choices of the one-form used for gauging, were obtained; selfdual gravity and selfdual supergravity are among these.
Perhaps the most intriguing one was the case of gauging with a weightless one-form which seemed to lead to $\N =8$ supergravity.
A twistor string descriptiuon of $\N =8$ supergravity will have implications for the finiteness of certain loop calculations \cite{bern} and would also explain the known and suggestive formulae for graviton scattering amplitudes \cite{nair}.
In this note we will focus on this special case and calculate some graviton scattering amplitudes; we will present evidence that this twistor string theory is actually $\N =8$ chiral supergravity and not $\N =8$ Einstein supergravity.

We begin with the basics of Berkovits' version of the twistor string theory \cite{berk}.
The target space variables are the supertwistors $Z^I = (\omega^A, \pi_{A'}, \psi^a)$,
$A, A' = 1,2$, $ a= 1, ...,4$, and $Y_I$ are the conjugate variables.
We also have ${\tilde Z}^I , {\tilde Y}_I$ with $Z^I = {\tilde Z}^I$, $Y_I = {\tilde Y}_I$
on the world-sheet boundary. For genus zero, the world-sheet may be taken as the upper hemisphere of a ${\bf CP}^1$. The action is given by
\beq
S = \int ~ {Y} (\bdel + \bA ) Z  ~+~ S_C ~+~{\rm tilde ~part}\label{1}
\eeq
Here $\bA$ is a $GL(1, {\bf C})$ gauge field and the action (\ref{1}) has the gauge symmetry,
\beq
Z \rightarrow \lambda Z , \hskip .1in Y \rightarrow  \lambda^{-1} Y, \hskip .1in 
\bA \rightarrow \bA - \bdel \log \lambda\label{2}
\eeq
This symmetry will help to reduce the physical set of variables to the projective
supertwistor space.

Consider first the integration over the fields $\bA$ in a functional integral with the action (\ref{1}).
On the two-sphere or ${\bf CP}^1$, there are topologically nontrivial $GL(1, {\bf C})$
configurations, corresponding to Dirac monopoles. 

The space of gauge potentials then splits up into a set of disconnected pieces, one for each value of the monopole number, say, $d$. In each sector, we can write
$\bA = \bA_d + \delta \bA$, where $\bA_d$ is a fixed configuration of monopole number
$d$ and $\delta \bA$ is a fluctuation (of zero monopole number).
In two dimensions, we can always write $\delta \bA = \bdel \Theta$ for some complex
function $\Theta$ on ${\bf CP}^1$. This means that configurations within the same connected component can be gauge transformed to $\bA_d$ by the $GL(1, {\bf C})$ gauge transformation with parameter $\Theta$. The volume element for the space of gauge potentials is
then given as
\beq
[d\bA ] = \sum_d ~[d\Theta ]~\det \bdel
\label{3}
\eeq
The invariant integration then reduces to a summation over $d$
with a factor $\det \bdel$. Using $\bA = \bA_d + \bdel \Theta$ in the action
(\ref{1}) and eliminating $\Theta$ by a $GL(1, {|bf C})$ transformation, we see that the correlation function of a set of vertex operators, $V_1 , ..., V_n$ is given by
\beqar
{\cal M}&=&\sum_d {\cal C}_d~ M_d\nonumber\\
M_d &=& \int (\det \bdel ) ~ e^{-S_C }~ e^{- \int Y (\bdel +\bA_d )Z }~ ~~V_1 V_2 ...V_n
\label{4}
\eeqar
 where ${\cal C}_d$ are some normalization constants.
Notice that, since we have equal numbers of charged fermionic and bosonic fields in
the action (\ref{1}), there is no anomaly for the $GL(1, {\bf C})$ transformation.
The factor $\det \bdel$ is a constant but it does contribute to the conformal
anomaly, raising the required central charge of the $S_C$-part of the action to $28$.

We now turn to the functional integration over the fields $Z^I, Y_I$. This is most conveniently done by expanding $Z^I$ in modes of $(\del +A ) (\bdel +\bA )$ and integrating over the mode coefficients. The key point is that there are zero modes of $(\bdel + \bA)$ in the expansion for $Z^I$,
\beqar
Z^\alpha &=& \sum_{\{a\}} a^\alpha_{a_1 a_2 \cdots a_d} ~ u^{a_1} u^{ a_2} \cdots u^{ a_d}
 ~+ \sum b^{\alpha, i_1...i_k}_{a_1 ...a_d j_1 ...j_k} \bu_{i_1}
\cdots\bu_{i_k}  u^{a_1} \cdots u^{ a_d} u^{j_1}\cdots u^{j_k}\nonumber\\
\psi^a &=& \sum_{\{ a\}} \gamma^a_{a_1 a_2 \cdots a_d} ~ u^{a_1} u^{ a_2} \cdots u^{ a_d} ~+ \sum  \zeta^{\alpha, i_1...i_k}_{a_1 ...a_d j_1 ...j_k} \bu_{i_1}
\cdots\bu_{i_k}  u^{a_1} \cdots u^{ a_d} u^{j_1}\cdots u^{j_k} \label{5}
\eeqar
The first set of terms on the right hand side correspond to zero modes of $(\bdel +\bA)$; these are holomorphic in the $u$'s. These terms give a holomorphic map of ${\bf CP}^1$
in twistor space, hence a holomorphic curve of degree $d$. 
$\{ a \}, \{ \gamma \}$ are the moduli of this holomorphic curve
in supertwistor space.
The higher terms on the right hand side of
(\ref{5}) correspond to nonzero modes of $(\del +A ) (\bdel +\bA )$ and have $\bu$'s with
$N_u = N_{\bu} = d$ where $N_u, N_{\bu}$ are the numbers of $u$'s and $\bu$'s, respectively. The $Y_I$'s have a mode expansion with the mode functions being conjugates of the nonzero modes used for $Z^I$,
\beqar
Y_\alpha &=& 
\sum {\bar b}^{i_1 ...i_k}_{\alpha, a_1 ...a_d j_1 ...j_k} \bu_{i_1}
\cdots\bu_{i_k}  u^{a_1} \cdots u^{ a_d} u^{j_1}\cdots u^{j_k}\nonumber\\
Y_a &=&  \sum {\bar\zeta}^{i_1...i_k}_{a,  a_1 ...a_d j_1 ...j_k} \bu_{i_1}
\cdots\bu_{i_k}  u^{a_1} \cdots u^{ a_d} u^{j_1}\cdots u^{j_k} \label{6}
\eeqar
The $Y$'s do not have a zero mode part. The integration over $(b, {\bar b})$, $(\zeta, {\bar\zeta})$ leads to Wick contractions of $Y$, $Z$ in any correlation function.
This means that the nonzero mode part of the $Z$'s have to be paired (with the $Y$'s)
to get a nonzero contribution. Thus, once all the $Y$'s in the integrand have been Wick-contracted with the $Z$'s, the nonzero mode parts of the remaining $Z$'s give zero.
We may therefore replace the remaining $Z$'s by their zero mode part. This leads to a simple rule for calculating correlators:
\begin{quotation}
\noindent For any correlator, carry out $Y$-$Z$ Wick contractions (integrate over mode coefficients of
nonzero modes), then replace
all $Z$'s by zero modes. Integrate over the coefficients $\{ a\}, \{ \gamma \}$,
(moduli of holomorphic curves) to obtain the correlator.
\end{quotation}

It is interesting to see how this works out for the gauge field scattering amplitudes.
In this case, the vertex operator is given by $V = \int d\sigma \Phi_{p,\eta} (Z) ~J(\sigma )$, where $J(\sigma )$ is the $GL(1, {\bf C})$ current from the $S_C$-part of the action and
\beq
\Phi_{p,\eta}(Z) = \delta (\pi\cdot \bp ) \left( {\pi \cdot \alpha \over \bp \cdot \alpha}
\right) ~\exp\left[ (\omega\cdot p + \psi \cdot \eta ) {\bp \cdot \alpha \over \pi \cdot \alpha}
\right]
\label{7}
\eeq
$\Phi_{p,\eta}(Z)$ is holomorphic in $Z$ of degree zero and it is of degree $-2$ in the spinor momentum $\bp_{A'}$. The Grassmann variable $\eta^a$ characterizes the helicity state, the term with $k$ $\eta$'s corresponding to the state of helicity $1- \half k$.
Since there are no $Y$'s in $\Phi$, there are no Wick contractions to be done. In the correlator of $V$'s, we may, therefore, replace $Z$'s by their zero mode part
\beqar
Z^\alpha_{cl} &=& \sum_{\{a\}} a^\alpha_{a_1 a_2 \cdots a_d} ~ u^{a_1} u^{ a_2} \cdots u^{ a_d}
\nonumber\\
\psi^a_{cl} &=& \sum_{\{ a\}} \gamma^a_{a_1 a_2 \cdots a_d} ~ u^{a_1} u^{ a_2} \cdots u^{ a_d}  \label{8}
\eeqar
Notice that these are solutions of the classical equation of motion 
$\bD Z\equiv (\bdel +\bA)  Z =0$. We have indicated this by the subscript on the fields
$Z^\alpha, \psi^a$.
From the world-sheet point of view the scattering amplitude for gauge particles is entirely classical. Thus the amplitude can be calculated by solving the classical equation of motion
$\bD Z =0$ and using these in the product $V_1 V_2 \cdots V_n$.
Since there are many classical solutions parametrized by $\{ a, \gamma \}$, we must integrate over the moduli space of classical solutions. The overall $GL(1, {\bf C})$ scale invariance implies that one complex scale factor can be removed from the set $\{ a, \gamma \}$.
Further, since the solutions are holomorphic in the $u$'s, physical quantities are invariant under the $SL(2, {\bf C})$ transformation $u^a \rightarrow {u^a}~' = g^a_{~b} u^b$,
$g\in SL(2, {\bf C})$.
This is equivalently expressed in terms of $\{ a, \gamma \}$ since
$Z^I( a, \gamma , gu ) = Z^I (a', \gamma', u)$
with
\beqar
a'^\alpha_{a_1 a_2 \cdots a_d} &=& a^\alpha_{b_1 b_2 \cdots b_d}~ g^{b_1}_{~a_1}
g^{b_2}_{~a_2}\cdots g^{b_d}_{~a_d}\nonumber\\
\gamma'^\alpha_{a_1 a_2 \cdots a_d} &=& \gamma^\alpha_{b_1 b_2 \cdots b_d}~ g^{b_1}_{~a_1}
g^{b_2}_{~a_2}\cdots g^{b_d}_{~a_d}\label{9}
\eeqar
Thus wecan remove $SL(2, {\bf C})$-worth of parameters from $\{ a, \gamma \}$.
With the scale factor, this is equivalent to a $GL(2, {\bf C})$ symmetry, so that the measure on the moduli space of classical solutions (for the degree $d$ holomorphic curve) is
\beq
d\mu_d = {d^{4d+4}a~~ ~d^{4d+4} \gamma \over 
vol [GL(2,{\bf C})]}
\label{10}
\eeq
The scattering amplitude for the degree $d$ curve is given by
\beqar
\la V_1 V_2 \cdots V_n \ra_d &=& \int d\mu_d ~\int d\sigma_1 d\sigma_2
\cdots d\sigma_n ~~\left[ \Phi_{p_1, \eta_1 } (Z_{cl})
\Phi_{p_2, \eta_2 } (Z_{cl})\cdots \Phi_{p_n, \eta_n } (Z_{cl})\right]~\nonumber\\
&&\hskip 1in \times
\la J(\sigma_1) J(\sigma_2) \cdots J(\sigma_n) \ra 
\label{11}
\eeqar

In the Witten-Berkovits string theory, there are also vertex operators corresponding to conformal supergravitons. We will not need them at this stage.

The modification of the action corresponding to the new set of twistor string theories is given by \cite{AHM}
\beq
S =\int [ Y \bD Z ~+~ {\bar B}_i K_i ] ~+~ S_C ~+~ {\rm tilde~ part}
\label{12}
\eeq
where $K_i$ is a holomorphic one-form on twistor space; it is of the form
$K_i = k_{i I } \del Z^I$.
The action (\ref{12}) has a new gauge invariance given by the transformation
\beq
\delta {\bar B}_i = \bdel \Lambda_i, \hskip .2in \delta Z^I =0, \hskip .2in \delta Y_I = k_{iI }\del \Lambda_i + 2\Lambda_i 
k_{i[I,J]} \del Z^J\label{13}
\eeq
The one-forms $K_i$ can, in general, be taken to carry $GL(1, {\bf C})$ charge $h_i$.
Also, in general, one can take the number of fermionic components $\psi^a$ to be $\N$, rather than $4$. The conformal and $GL(1, {\bf C})$ anomalies are then given by
\beqar
C&=& 4 -\N - 28 + C_C - 2 (\delta -n)\nonumber\\
\kappa &=& 4 -\N - \sum_i \epsilon_i (h_i )^2
\label{14}
\eeqar
Here $\delta $ is the number of bosonic one-forms and $n$ is the number of fermionic one-forms among the chosen set of $K_i$'s. Further, $\epsilon_i = 1$ for bosonic $K_i$ and $\epsilon_i = -1$ for
fermionic $K_i$. With these formulae, one can see that there are many anomaly-free solutions, giving many new string theories. 
For example, the choice $\N =0$ and $K = \epsilon^{A'B'} \pi_{A'} d \pi_{B'}$ leads to selfdual gravity (with no supersymmetry).
Many other cases are given by Abou-Zeid, Hull and Mason in reference \cite{AHM}.
Perhaps the most interesting case corresponds to the $\N =4$ (so that we are back to the
$\N =4$ supertwistor space) with
\beq
K = w(\pi )~ \epsilon^{A'B'} \pi_{A'} d \pi_{B'}
\label{15}
\eeq
where $w(\pi )$ is of degree $-2$. The gauging is thus done with a weightless one-form.
In this case $h =0$ and $\delta = 1, n=0$.
This seems to lead to $\N =8$ supergravity; the field content of the theory is expressed in terms of $\N=4$ multiplets.

There are two types of vertex operators for gravitons,
\beq
V_f = \int  f^I Y_I, \hskip .3in V_g = \int  g_I \del Z^I\label{16}
\eeq
For these operators to be physical, the functions $f^I, g_I$ must obey the constraints
\beq
\del_I f^I =0, \hskip .3in g_I Z^I =0
\label{17}
\eeq
There is also an equivalence relation $f^I \sim f^I +\delta f^I$,
$g_I \sim g_I +\delta g_I$, for functions differing by the transformations,
\beq
\delta f^I = \lambda ~Z^I, \hskip .3in \delta g_I = \del_I \chi
\label{18}
\eeq
for arbitrary $\lambda , \chi$. The requirements (\ref{17}) and (\ref{18})  are the same as in the original
Witten-Berkovits theory and lead to conformal supergravity multiplet.
The new gauge symmetry leads to the further condition
\beq
\epsilon^{A' B'} f_{A'} \pi_{B'} =0 
\label{19}
\eeq
and the equivalence relation $ g^{A'} \sim g^{A'} +\delta g^{A'}$,
\beq
\delta g^{A'} = \epsilon^{A'B'} \pi_{B'} ~\xi
\label{20}
\eeq
for some function $\xi$. The vertex operator which obeys these conditions may be obtained as
\beq
V_f = \int d\sigma ~\epsilon^{AB}  {\del h \over \del \omega^B} ~Y_A ~+~ \int d\sigma f^a Y_a
\label{21}
\eeq
The function $h$ is of degree of homogeneity $2$,
\beq
h = h_2 +  h^a_{ 3/2} \psi^a ~+\cdots + h_0 \psi^1 \psi^2 \psi^3 \psi^4
\label{22}
\eeq
This gives an $\N =4$ graviton multiplet starting with the graviton of helicity $+2$, represented
by $h_2$. The $f^a Y_a$-term will correspond to an $\N=4$ gravitino multiplet
starting with the ${3\over 2}$-helicity state.

For $V_g$, we may choose a representative $g_A$ of the form $g_A = \omega_A g$, because of the equivalence relation (\ref{18}). Further, the new equivalence relation (\ref{20}) tells us that $g^{A'}  = \tau^{A'}  {\tilde g} $ where $\tau \cdot \pi \neq 0$.
The constraint $g_I Z^I =0$ then implies that ${\tilde g}$ is determined by $g_a \psi^a$. Thus
the physical degrees of freedom are contained in $g$, which has degree of homogeneity $-2$, and in $g_a$, which is of degree $-1$.
We can write
\beq
g = g_0 ~+~ g^a_{-1/2} \psi^a ~+ \cdots + g_{-2}  \psi^1 \psi^2 \psi^3 \psi^4
\label{23}
\eeq
This describes an $\N =4$ negative helicity graviton multiplet. Likewise, $g_a$
will describe a negative helicity gravitino multiplet. There are also vertex operators for vector multiplets. All of these together can make up the $\N =8$ supergravity multiplet.

The explicit form of the vertex operators, particularly the graviton part, is useful for explicit calculations. They can be taken as
\beqar
h_2 &=& \delta ( \pi \cdot \bp ) \left( {\pi \cdot \alpha \over \bp \cdot \alpha }\right)^3
~\exp\left[ (\omega\cdot p + \psi \cdot \eta ) {\bp \cdot \alpha \over \pi \cdot \alpha}
\right]\nonumber\\
f^A &=& \epsilon^{AB} p_B {\bp \cdot \alpha \over \pi \cdot \alpha}~h_2
\equiv \epsilon^{AB} K_B~h_2
\label{23a}\\
g_{-2} &=&  \delta ( \pi \cdot \bp ) \left( {\bp \cdot \alpha\over \pi \cdot \alpha }\right)^5
~\exp\left[ \omega\cdot p~ {\bp \cdot \alpha \over \pi \cdot \alpha}
\right]
\label{23b}
\eeqar

The three-graviton amplitude for the choice of helicities $(++-)$ was calculated by AHM in their original paper. This is given by the correlator $\la V_f (1) V_g (2) V_g (3) \ra$ and is nonzero for curves of degree zero. Explicitly,
\beq
\la V_f(1) V_f(2) V_g(3)\ra_{d=0} =
\left( {\la 12\ra^3 \over \la 31\ra \la 23\ra }\right)^2 ~ \delta^{(4)}(p_1 +p_2 +p_3)
\label{24}
\eeq
This result is in agreement with Einstein supergravity. While this is an encouraging result, it is necessary to calculate amplitudes for other helicities, at least the $(+--)$-amplitude, to see whether the theory is truly Einstein supergravity or some chiral version of it.
We shall now calculate the latter amplitude.
This is contained in the correlator $\la V_f (1) V_g (2) V_g (3) \ra$.
This amplitude cannot get contributions from curves of degree zero, but can have nonzero
contributions from curves of degree one.
In a direct calculation of this amplitude, we will encounter only one Wick contraction $\la Y Z \ra$ since there is only one factor of $V_f$. Therefore, this amplitude, considered as a correlator in the world-sheet field theory, will be entirely classical.
We can calculate it by solving the equations of motion for the action (\ref{12}) with vertex operator sources,
\beq
S = \int [ Y \bD Z ~+~ {\bar B}_i K_i ] ~+~ S_C ~+~ \oint \lambda g_I \del Z^I + \oint \rho f^I Y_I ~+~ {\rm tilde~ part}
\label{25}
\eeq
The equation of motion for $Z^I$ is given by
\beq
(\bD Z)^I ~+~ \rho ~ f^I \delta (  \vert \sigma \vert - \sigma_0 ) = 0
\label{26}
\eeq
The $\delta$-function localizes the source term on the boundary.
For $\bA$'s of degree one, the solution to the equation (\ref{26}) is of the form
\beq
Z^I = a^I_\alpha ~u^\alpha ~-~ \int_2 \left( {1\over \bD} \right)_{12} \left[\rho f^I (a^J_\alpha u^\alpha )\right]_2~+\cdots
\label{27}
\eeq
Evaluating the action on this solution, we find
\beqar
S &=& \oint \lambda~g_{I cl}\del Z^I_{cl}\nonumber\\
&=& \oint \lambda ~ g_I \del Z^I \Biggr]_{Z = au} ~-~ \int \lambda ~\rho~ {1\over 2}( \del_J g_I - \del_I g_J) \left( {1\over \bD}\right)_{12} f^J (Z(2)) \Biggr]_{Z =au} ~+\cdots
\label{28}
\eeqar
The scattering amplitude is given by integral of $e^{-S}$ over the moduli of solutions, and for $\la V_f V_g V_g \ra$, we are interested in the term with two powers of $\lambda$ and one power of $\rho$. Thus
\beq
\la V_f (1) V_g (2) V_g (3) \ra_{d=1}   = \int d\mu \oint f^I (1) \left( {1\over \bD}\right)_{12}{1\over 2} \bigl[ ( \del_I g_J - \del_J g_I )
\del Z^J\bigr]_2   ( g_K \del Z^K )_3
~+~ ( 2\leftrightarrow 3 )
\label{29}
\eeq
The remaining integration is over the moduli space of the classical solutions, namely, over
$a^I_\alpha$. For this purpose, we need to identify the physical set of classical solutions.
In the Witten-Berkovits string theory, the $GL(2, {\bf C})$ invariance allowed us to choose 
$\pi_{A'} = a_{A'\alpha} u^\alpha = u_{A'}$. In other words, we could set $a_{A'\alpha} = \delta_{A' \alpha}$, which led to $\omega^A = x^{AA'} u_{A'} = x^{AA'} \pi_{A'}$.
The moduli space of the line, for the bosonic part, is thus a copy of spacetime in twistor space.
In the present case, the situation is somewhat different.
The new gauge symmetry gives the constraint
\beq
w(Z)~\epsilon^{A'B'} \pi_{A'} \del \pi_{B'} =0
\label{30}
\eeq
This may be viewed as the equation of motion for ${\bar B}$. Using $\pi_{A'}= a_{A'\alpha }
u^\alpha$, and the fact that $w(Z)$ must be chosen to have singularities away from 
the world-sheet on ${\bf CP}^1$, this equation gives
\beq
\det a = \half \epsilon^{A'B'} \epsilon^{\alpha\beta}~a_{A'\alpha} a_{B' \beta}  = 0
\label{31}
\eeq
By $SL(2, {\bf C})$ invariance we can bring the solution set of this condition to the form
\beq
a_{A' \alpha } = \left( { \begin{matrix} 1 &0 \\
\zeta &0\\
 \end{matrix} }\right)
\label{32}
\eeq
In other words, $\pi_{1'} = u^1$, $\pi_{2'} = \zeta u^1$, where $\zeta$ is the same for all vertex operators.
Unlike the previous case, we do  not get a copy of spacetime in the twistor space.
It is clear that the amplitude (\ref{29}) will, therefore, vanish. This can be seen in detail as follows.
The $SL(2, {\bf C})$ transformation which brings $a_{A'\alpha}$ to the form
(\ref{32}) can be written as
\beq
g = \left( {\begin{matrix} {(1- a_{12}~ \chi )/a_{11}}& - a_{12}\\
\chi& a_{11}\\
\end{matrix} } \right)
\label{33}
\eeq
$\chi$ is still a free paramter and we see  the form (\ref{32}) is unchanged by
transformations of the form
\beq
g = \left( {\begin{matrix} 1& 0\\
\chi& 1\\
\end{matrix} } \right)
\label{34}
\eeq
For $u^\alpha$ this leads to the change $ u^1 \rightarrow u^1$, $u^2 \rightarrow
\chi u^1 + u^2$, so that $\sigma = u^2/u^1$ changes as $\sigma \rightarrow \sigma +\chi$.
Thus we can use this freedom to set one of the $\sigma$'s, say $\sigma_3$ to zero.
Further, we write $\omega^A = a^A_\alpha u^\alpha = u^1 ( b^a + \sigma~ {\tilde b}^A )$
and $\psi^a = \gamma^a_\alpha u^\alpha$. The integration over the fermionic variables
$\gamma^a_\alpha$ leads to a factor $(u_2\cdot u_3)^4\sim (\sigma_2 - \sigma_3 )^4$.
Putting all these together, the correlator in (\ref{29}) becomes
\beqar
\la V_f (1) V_g (2) V_g (3) \ra_{d=1} \!\!  &=& \!\!\int {[d\zeta db]\over GL(1,{\bf C})}
\delta (\pi \cdot \bp_1)\delta (\pi \cdot \bp_2)\delta (\pi \cdot \bp_3)
{( \bp_2\cdot \alpha )^5 (\bp_3 \cdot \alpha )^5 \over (\bp_1 \cdot \alpha )^3 (\pi \cdot \alpha )^7} \nonumber\\
&&\hskip .4in \times \exp \left( b^A [K_{1A} + K_{2A } + K_{3A } ] \right)\times  {\cal I}
\nonumber\\
&&\hskip .5in+~{( 2\leftrightarrow 3 )}
\label{35}
\eeqar
where ${\cal I}= {\cal I}_1 ~+~ {\cal I}_2$, with
\beqar
{\cal I}_1 &=& \left( 2 + b \cdot K_2  \right)\int [d\sigma_1 d \sigma_2 d{\tilde b}] ~{\sigma_2^4 \over (\sigma_1 - \sigma_2)}~ K_1 \cdot {\tilde b} ~b \cdot {\tilde b}
~e^{ \left( \sigma_1 {\tilde b}\cdot K_1 + \sigma_2 {\tilde b}\cdot K_2 \right)}\nonumber\\
{\cal I}_2 &=& \int [d\sigma_1 d \sigma_2 d{\tilde b}]~{\sigma_2^4 \over (\sigma_1 - \sigma_2)} ~ K_1 \cdot {\tilde b} ~b \cdot {\tilde b}
\left[  \sigma_2 {\tilde b} \cdot K_2 \right]
e^{ \left( \sigma_1 {\tilde b}\cdot K_1 + \sigma_2 {\tilde b}\cdot K_2 \right)}
\label{36}
\eeqar
Also, $K_A$ stands for $p_A ~\bp\cdot \alpha /\pi\cdot \alpha$.

Consider now the evaluation of ${\cal I}_1$. The factors of ${\tilde b}$ inside the integral can be replaced by derivatives with respect to $K_1$ of the exponential by defining
\beq
{\cal Q} = (2 + b \cdot K_2 ) ~ K_1 \cdot {\del \over \del K_1} ~b \cdot {\del \over \del K_1}
\label{37}
\eeq
We can then write, denoting $\sigma_1 - \sigma_2 = \sigma_{12}$,
\beqar
{\cal I}_1 &=& {\cal Q}  \int [d\sigma_1 d \sigma_2 d{\tilde b}] 
{\sigma_2^4 \over \sigma_{12} ~\sigma_1^2} ~e^{ \left( \sigma_1 {\tilde b}\cdot K_1 + \sigma_2 {\tilde b}\cdot K_2 \right)}\nonumber\\
&=&  {\cal Q}  \int [d\sigma_1 d \sigma_2] 
{\sigma_2^4 \over \sigma_{12} ~\sigma_1^2}~ \delta ( \sigma_1 K_{11} + \sigma_2 K_{21} )
~\delta ( \sigma_1 K_{12} + \sigma_2 K_{22} )\nonumber\\
&=& - {\cal Q} \int d\sigma_2 ~ \sigma_2 {K_{11}^2 \over (K_{11} +K_{21} )}
\delta \left( \sigma_2 [ K_{22} - K_{21} K_{12} /K_{11} ] \right)\nonumber\\
&=& - {\cal Q} \int {K_{11}^3 \over (K_{11} +K_{21} )} {1\over K_1\cdot K_2}~
d\sigma_2~\sigma_2 ~\delta (\sigma_2) ~=~0
\label{38}
\eeqar
We have used a continuation to real values for the $\sigma$-integrals, as is done
for the twistor amplitudes for the gauge particles.
In ${\cal I}_2$ we have another factor with ${\tilde b}$ which gives an additional
$1/\sigma_1$; but there is also a factor of $\sigma_2$ multiplying it, so that the
powers of $\sigma_2$ in the last integral do not change and we get
${\cal I}_2 =0$ as well.
The scattering amplitude thus vanishes.

The theory we are discussing has the particle content of ${\cal N}=8$ supergravity.
The vanishing of the $(+ - -)$-amplitude is inconsistent with Einstein supergravity, rather it is suggestive of the theory being a chiral version of ${\cal N}=8$ supergravity, rather like the
theory introduced by Siegel \cite{siegel}.

\vskip .1in
I thank Mohab Abou-Zeid, Chris Hull and Lionel Mason for discussions.
This research was supported in part by the National Science Foundation 
grant PHY-0555620 and by a PSC-CUNY grant.
\fontfamily{cmr}\fontsize{11pt}{16.7pt}\selectfont

\end{document}